# Object-Oriented Program Comprehension: Effect of Expertise, Task and Phase

**Jean-Marie Burkhardt [1, 2], Françoise Détienne [1]
and
Susan Wiedenbeck [3]**

[1] Eiffel Group, "Cognition and Cooperation in Design"
INRIA
Domaine de Voluceau, Rocquencourt, BP 105,
78153, Le Chesnay, cedex, France
Jean-Marie.Burkhardt@inria.fr, Francoise.Detienne@inria.fr

[2] Université Paris 5 René Descartes
45 rue des Saint Pères
75005 Paris, France

[3] College of Information Science and Technology
Drexel University
Philadelphia, PA 19104, USA
susan.wiedenbeck@drexel.edu

**ABSTRACT** The goal of our study is to evaluate the effect on program comprehension of three factors that have not previously been studied in a single experiment. These factors are programmer expertise (expert *vs.* novice), programming task (documentation *vs.* reuse), and the development of understanding over time (phase 1 *vs.* phase 2). This study is carried out in the context of the mental model approach to comprehension based on van Dijk and Kintsch's model (1983). One key aspect of this model is the distinction between two kinds of representation the reader might construct from a text: 1) the textbase, which refers to what is said in the text and how it is said, and 2) the situation model, which represents the situation referred to by the text. We have evaluated the effect of the three factors mentioned above on the development of both the textbase (or program model) and the situation model in object-oriented program comprehension. We found a four-way interaction of expertise, phase, task and type of model. For the documentation group we found that experts and novices differ in the elaboration of their situation model but not their program model. There was no interaction of expertise with phase and type of model in the documentation group. For the reuse group, there was a three-way interaction between phase, expertise and type of model. For the novice reuse group, the effect of the phase was to increase the construction of the situation model but not the program model. With respect to the task, our results show that novices do not spontaneously construct a strong situation model but are able to do so if the task demands it.

# 1. Objectives

In recent years a body of empirical studies on computer program comprehension has developed. This research differs in the theoretical approach taken and the factors included in the empirical studies. However, one factor that has been relatively constant is that most studies have been carried out in the procedural paradigm. While the procedural programming paradigm remains widespread, the object-oriented (OO) paradigm has taken on increasing prominence in industry and education. It is important to extend the study of program comprehension to the OO paradigm, because comprehension is such a ubiquitous task. Any features of the OO paradigm that affect comprehension will have an effect on the performance of comprehension-related tasks, such as program debugging, maintenance, and reuse.

In this research we present a model of OO program understanding. We then report on an empirical study of program understanding in the OO paradigm. The goal of our study is to evaluate the effect on program comprehension of three factors that have not previously been studied in a single experiment. These factors are programmer expertise, programming task, and the development of understanding over time. This study is carried out in the context of the mental model approach to comprehension. Based on van Dijk and Kintsch's model of text understanding (1983), Pennington (1987a, 1987b) developed and tested a model of procedural program comprehension. One key aspect of van Dijk and Kintsch's model is the distinction between two kinds of representation the reader might construct from a text: 1) the textbase, which refers to what is said in the text and how it is said, and 2) the situation model, which represents the situation referred to by the text. We will evaluate the effect of the three factors mentioned above on the development of both the textbase and the situation model in object-oriented program comprehension.

The organization of the rest of the paper is as follows. Section 2 describes the mental model approach to procedural program comprehension and the effect of expertise and task in procedural program comprehension. Section 3 presents a framework for studying OO program understanding, then summarizes the few existing studies on the role of expertise and task in OO program comprehension. This is followed in section 4 by our research questions. In section 5, we describe the methodology used in our empirical study. In section 6 we present the results of this study, and in section 7 we discuss the results. Section 8 presents directions for future research.

# 2. The mental model approach to procedural program comprehension

## 2.1 Description of the mental model approach

The mental model approach to program comprehension is based on the recent evolution of van Dijk and Kintsch's model of text comprehension (Kintsch, 1988; van Dijk & Kintsch, 1983), which takes into account the role of domain knowledge in text understanding. Three distinct, but interacting, levels of cognitive representation[1], are distinguished:

> level 1. the surface form representation,
> level 2. the propositional textbase representation,
> level 3. the situation model, or mental model.

---

[1] The rest of this paper will not deal with level 1.

Levels 1 and 2 are linguistic representations of the text. Level 1 reflects what is contained in the text at a surface, or verbatim, level. Level 2 is isomorphic with the text structure and reflects what is contained in the text at a propositional level, i.e., it represents the microstructure and the macrostructure of the text. Level 3 corresponds to an a-linguistic representation of the text and reflects the world situation referred to by the text. It is isomorphic or homomorphic with the situation described by the text. It is initially built up from a linguistic representation and makes extensive use of the reader's existing domain knowledge. It is produced by inferences and is also a source for making new inferences (Kintsch, 1988; Johnson-Laird, 1983; Schmalhofer and Glavanov, 1986). Text comprehension theory assumes that the propositional representation is built by means of automatic processes from the verbatim representation. The building of the situation model, by contrast, is optional (Mills, Diehl, Birkmire, and Mou 1995).

Pennington (1987a, 1987b) adapted van Dijk and Kintsch's text comprehension model to procedural program comprehension. Pennington distinguished between two different mental representations which may be built while comprehending a program: (1) the domain model which is equivalent to van Dijk and Kintsch's situation model and reflects entities of the problem domain and their relationships, and (2) the program model which is equivalent to van Dijk and Kintsch's propositional textbase and reflects the text-based representation of the program. Pennington argued that control flow and elementary operations information represent dynamic procedural episodes in a program and thus belong to the program model. On the other hand, program goals and data flow information together make up the representation of program function and belong to the domain, or situation, model. A summary of the correspondence that Pennington proposes between the program text, knowledge structures, and mental representations is shown in Table 1 (adapted from Pennington, 1987b). In this table, text relations refer to abstractions of the program text. Knowledge structures refer to relatively generic knowledge stored in long-term memory that must be activated to be used. Mental representation refers to the content of working memory at a particular point in the comprehension activity, constructed from activated knowledge in long-term memory, the results of prior comprehension episodes, and external information gathered from the environment. Model refers to the program model or situation model, as described by Pennington.

INSERT TABLE 1 ABOUT HERE

Elementary operations are the simple operations making up a program, usually corresponding to a single line of the program text. Thus, they may be considered analogous to propositions in natural language texts. Elementary operations provide a dynamic view of a program in that they are executable, action-oriented program elements that lead to changes of state. They may also be considered functional in that they correspond to program subgoals at a very low-level of granularity. Control flow concerns the order in which elementary operations are carried out. Possible control flow methods are sequential, branching based on a condition, and looping. Since control flow defines the relationship of elementary operations during execution, it is analogous to the links between propositions in text understanding. A control flow perspective gives a dynamic view of a program because control flow instructions cause actions that transform the current state of the program. Both elementary operation and control flow information are explicitly present in a program text.

According to Rist (1996), the deep structure of a program is made up of plans, i.e., methods of structuring programming actions to achieve a desired goal, and the objects that participate in plans. Knowledge of main program goals in Pennington's model corresponds to knowledge about the goals of plans. It is part of the situation model because it represents knowledge about the outcome of programming plans in terms of the situation they represent in the world. Data flow concerns the transformations that objects undergo in the course of plan execution. Data flow may be considered dynamic in that it concerns transformations in the program. However, it may also be considered functional in that the goals of a plan are achieved through the act of modifying data. Data flow is part of the situation model because it represents the changes made to the world situation represented in a program. The main goals must be inferred in a program. Data flow, on the other hand, is explicitly represented in a program text; however, it is generally not highlighted by the structure of the program as is control flow and, as a result, may be difficult to extract.

In testing her model, Pennington's experimental paradigm was to give participants a program to read for a limited time and then ask them questions reflecting different information categories presumed to make up the program and situation models. The correctness of the responses served as an indicator of the nature of their representation. Her experiments (Pennington, 1987a; 1987b) supported the dual model. Control flow and elementary operations knowledge, which make up the program model, emerged earlier during program comprehension, perhaps because the programmer begins by reading elementary operations and grouping them into larger text structure units corresponding to the control flow. Program goals and data flow knowledge, which make up the situation model, emerged later with continued processing of the program.

Later work by Bergantz and Hassell (1991) verified Pennington's dual model for declarative program comprehension. Through the analysis of comprehension protocols, they found that in studying a PROLOG program experienced declarative programmers first concentrate their attention on program model information, including control flow and data structures. Later in the study of the program their attention shifts to information about program goals, as they build a situation model. Corritore and Wiedenbeck (1991) found that novice procedural programmers form a strong program model during study of a program, but a very weak situation model.

## 2.2 The effect of expertise in the mental model approach to program understanding

An important issue in program understanding is the effect of expertise on comprehension. Studies of text comprehension have shown that expertise in the domain described in a text aids comprehension (e.g., Bransford and Johnson, 1972; Moravcsik and Kintsch, 1993; Voss, Vesonder, and Spilich, 1980; Tardieu, Ehrlich, and Gyselinck, 1992). With respect to the role of expertise in the formation of the mental representation of programs, the mental model studies mentioned above all concern either experts or novices but do not compare the two. However, the fact that expert programmers in Bergantz and Hassell's (1991) study showed evidence of developing a situation model over time, while the novices in Corritore and Wiedenbeck's (1991) study did not, suggests a difference. Pennington (1987a) also suggests a difference in the mental representations of professional programmers of different levels of proficiency. Furthermore, some studies of program understanding based on a functional approach of schema activation and instantiation have clearly shown differences based on domain knowledge (Soloway and Ehrlich, 1984; Soloway, Ehrlich, Bonar, and Greenspan, 1982; Wiedenbeck, 1986). For example, Soloway and Ehrlich (1984) showed that programs written in an unplan-like way are difficult to understand, particularly for more experienced programmers, because they fail to evoke the programmers' plan knowledge.

## 2.3 The effect of task in the mental model approach to program understanding

The recent evolution of the mental model approach takes into account the effect of purpose for reading on text understanding. This literature suggests that the construction of the mental representation may be strongly affected by the purpose for reading (Mannes, 1988; Mills et al., 1995; Richard, 1990; Schmalhofer and Glavanov, 1986). These authors identify two main categories of purpose for reading: read-to-recall, in which a reader's task is to remember information from the text, for example to paraphrase it, and read-to-do, in which the reader must carry out an activity that makes use of information in the text, for example carry out a procedure described in the text. It appears that a read-to-recall purpose focuses the understanding activity on the construction of the textbase (what is said and how it is said), whereas the read-to-do purpose focuses the understanding activity on the construction of the situation model. For example, Mills et al. found that individuals given read-to-recall instructions recalled a procedural text better, whereas individuals given read-to-do instructions performed the task described by the text better. Similarly, Schmalhofer and Glavanov found that individuals who studied for text summarization remembered more propositional information, while participants with the goal to carry out an activity using the information in the text remembered more situation information.

The effect of purpose for reading a program has not been explored systematically in the framework of the mental model approach to understanding. Nevertheless, studies have been carried out employing a variety of programming tasks and they may be discussed in terms of the read-to-recall vs. read-to-do distinction (Détienne, 1996). According to this interpretation, reading a program to document it can be considered similar to a read-to-recall task because it involves text summarization. As in text summarization, programmers who read a program to document it are expected to concentrate on encoding the program text itself, i.e., constructing a textbase representation. On the other hand, modifying a program or reusing code in design of a new program can be considered similar to a read-to-do task because it involves performing an activity that makes use of information in the program. Programmers who read a program to modify or reuse it are expected to concentrate on understanding the program in terms of its situation of reference.

### 2.3.1   Read-to-recall

The task of documenting a program produced by someone else has been studied by Rouet, Deleuze-Dordron and Bisseret (1995). Their hypothesis was that comments reflect the designer's cognitive representation of the entity being commented, which varies depending on the designer's expertise and the context of the documentation task. A posteriori analysis of comments showed that experts most often commented control flow and function at a low level, i.e., the function of a single line, equivalent to Pennington's elementary operations category. They did not comment the function of larger units of code. This suggests that programmers constructed a textbase representation. Another result was that "structural" units, e.g., beginnings of loops, were the most frequently commented. This suggests that the structure of the representation constructed reflects the structure of the program text as defined by the control structure.

Another study on documentation (Riecken, Koenemann-Belliveau & Robertson, 1991) showed that expert programmers documenting a program produced by someone else generated two times more concrete comments paraphrasing or explaining individual instructions than abstract comments communicating general domain information associated with the task of the program. These results suggest again that programmers in a documentation task construct a textbase representation rather than a situation model. Furthermore, participants added vertical spacing according to the program's text structure, e.g., between routines. This result also supports the textbase construction hypothesis, since participants used spacing in a manner that both preserved and emphasized text structure.

### 2.3.2 Read-to-do

The effect of the read-to-do purpose for reading may be examined in various tasks, e.g., program modification and reuse. In these situations programmers read a program in order to use it for performing a task. In program modification, the task is to change a program to meet new constraints or goals of the problem. In reuse, the task is to design or implement a target software program using parts of a source program. Results from studies of program modification and program reuse suggest that, as in text understanding studies, the read-to-do task has an effect on the encoding processes and entails the construction of a situation model.

Studies of the modification task (Littman, Pinto, Letovsky & Soloway, 1986; Koenemann & Robertson, 1991) show that differential encoding processes are involved. It was found that programmers use "as-needed" strategies. That is, they study code or documentation only if they believe that the code is relevant for the task. Koenemann and Robertson (1991) distinguish between three levels of relevance: direct relevance, i.e., code segments that have to be modified, intermediate relevance, i.e., code segments that are perceived to interact with relevant code, and strategic relevance, i.e., code which helps to locate or detect code of direct or intermediate relevance. As noted by the authors (p 129): "the [modification] task on hand determines the scope and focus of attention. For one modification it might be sufficient to know how a piece of code works, while for a different modification the question of why this implementation was chosen is of great importance."

Several studies of software reuse (Burkhardt and Détienne, 1995; Rouet, Deleuze-Dordron and Bisseret, 1995) show that, when a source component is evoked or retrieved in a problem solving phase (as opposed to an implementation phase) of software design, information about the source situation from which the component comes is searched for or inferred. Programmers infer solution goal structure, constraints, evaluation criteria, or design rationales. Thus, it seems that reusing a component implies more than constructing a textbase representation of the source component itself. It implies constructing a situation model of the source situation. This situation model allows the representation constructed for solving the design problem at hand to be enriched and the search space to be enlarged.

## 2.4 Interaction between expertise and task in program understanding

Although both the domain knowledge of the reader and the purpose for reading appear to affect program understanding, to our knowledge the interaction of these two factors has not been studied in the framework of the mental model approach. This appears to be a question worth investigating given the adaptability of many human cognitive processes. For example, programmers with low domain knowledge have been shown to fail to develop a strong situation model. However, they might be capable of building a situation model if they were given a task that required situation knowledge.

## 3. Applying the mental model approach to object-oriented program comprehension

## 3.1 Proposed model of comprehension of object-oriented programs

Our objective is to evaluate the effects of expertise, task, and their interaction in object-oriented program understanding. In pursuit of that objective, we have developed a model of OO program comprehension based on the mental model approach (see Table 2). This model is founded on Pennington's model (1987a, 1987b) in defining a program model and a situation model. It expands Pennington's model to take into account additional factors: object-oriented program features such as objects and message passing, the structure of larger programs, and the task.

INSERT TABLE 2 ABOUT HERE

As in Pennington's work, the program model contains text structure knowledge. Pennington defined text structure knowledge in terms of the microstructure of the program text: elementary operations, generally consisting of single lines of code, and control flow between these operations. However, since she was working with small programs, she did not consider the text macrostructure, i.e. the representation of larger text units such as routines. Nevertheless, these larger text units are important structural units in all but the smallest programs (Wiedenbeck, Fix & Scholtz, 1993). In our work, text microstructure consists of basic units of text and their links, while text macrostructure consists of larger units of text, as defined below:

- elementary operations. Elementary operations, forming part of the text microstructure, constitute basic text units usually consisting of one or a few lines of code.

- control flow. Also forming part of the text microstructure, control flow constitutes the links between text units. Control flow, at this fine level of granularity, represents the control structure (either sequence, loop or test) linking individual operations within a routine.

- elementary functions. Elementary functions consist of larger units of text, and thus form part of the text macrostructure. These functions correspond to units in the program structure, i.e., routines attached to objects.

Elementary operations, control flow, and elementary functions are all procedural in nature and may be characterized as dynamic representations because they provide a view of the program as an executable text. In addition, elementary operations and elementary functions can be considered functional, as well, because they provide a view of program function at a low level of granularity.

Our situation model contains problem knowledge and plan knowledge. Pennington considered the situation model to consist of the main goals and data flow relations in the program text. Main goals corresponded to the function of the program and data flow to the links between variables in a local plan unit within a routine. In our model, we supplement these relations to take into account the OO nature of programs and program size.

Pennington conducted her experiments with procedural languages and she did not examine representations of objects or even of data structures. However in OO programs, objects are central entities that map to the problem domain (Rosson and Alpert, 1990; Lee and Pennington, 1994), and the construction of the representation of objects should be taken into account in a model of OO program understanding (Burkhardt, 1997; Burkhardt, Détienne, and Wiedenbeck, 1997). In the current work, we assume that the representation of objects is part of the situation model inasmuch as it reflects the objects of the problem situation and the relationship of objects (Lee and Pennington, 1994).

In terms of program size, Pennington's model accounts for understanding of short programs but does not scale up easily to larger programs because it does not account for the representation of delocalized plans. Pennington assumes that the reader uses plan knowledge to construct the situation model. A plan is a set of actions that, when placed in a correct order, achieves some desired goal. Programmers have knowledge about patterns of program instructions which typically are used to accomplish certain functions or goals (Soloway, Ehrlich and Bonar, 1982). Pennington assumes that plan representations of a program are primarily based on data flow relations. In long programs, particularly in OO programs, many plans are delocalized. According to Rist (1996), plans and objects are orthogonal in OO systems. A plan can use many objects and an object can be used in many plans. In an OO system, the actions in a plan are encapsulated in a set of routines, and the routines are divided among a set of classes and connected by control flow. In our model we take the view that the construction of these complex delocalized plan representations is primarily based on client-server relationships, in which one object processes and supplies data needed by another object.

As a result of these considerations, we expand the situation model to include objects and their relationships, as well as client-server relationships of objects, as defined below:

- problem objects. These objects directly model objects of the problem domain.

- relationships between problem objects. These consist of the inheritance and composition relationships between objects.

- computing or reified objects. An example of a computing, or reified, object is a string class, which is not a problem domain object per se. Reified objects are represented at the situation model level inasmuch as they are necessary to complete the representation of the relationships between problem objects, i.e., they bundle together program-level elements needed by the domain objects.

- main goals. The main goals of the problem correspond to functions accomplished by the program viewed at a high level of granularity. They do not correspond to single program units. Rather, the complex plan which realizes a single goal is usually a delocalized plan in an OO program.

- client-server relationships. Communication between objects corresponds to client-server relationships in which one object processes and supplies data needed by another object. These connections between objects are the links connecting units of complex delocalized plans. In an OO system, the actions in a complex plan which performs a main goal are encapsulated in a set of routines, and the routines are divided among a set of classes and connected by control flow. Client-server relationships represent those *delocalized* connections.

- data flow relationships. Communication between variables corresponds to data flow relationships connecting units of *local plans within a routine.*

In the current model, objects and relationships between objects may be considered to form an object view of the program. The main goals of the program provide a functional view at a high level of granularity. Client-server and data flow relations provide a dynamic view in that they correspond to actions resulting in transformations of variables.

## 3.2  The effect of expertise and task in OO program understanding

A few studies have been carried out of OO program comprehension within the mental model approach (Wiedenbeck, Ramalingam, Sarasamma, and Corritore, 1999; Corritore and Wiedenbeck, 1999). However, these studies have either focused on novices and very small programs exclusively or on the comparison of comprehension in the OO and procedural paradigm. Comparing student and professional programmers, Boehm-Davis, Holt and Schultz (1992) found an effect of program design (in-line, functional or OO) on modification time. As the various versions of the programs were simulated in Pascal, the OO version lacked key characteristics of inheritance and polymorphism, which make the results difficult to generalize to the OO paradigm. To our knowledge, there are no existing studies comparing OO novices and experts using the mental model approach. This is a question worth pursuing because evidence about procedural program comprehension, though very limited, suggests that programmers of different levels of expertise may form different mental representations of a program (see section 2.2).

With respect to the effect of task in OO program comprehension, there is also a lack of empirical evidence. Existing studies of procedural or OO program understanding (Pennington, 1987a; Corritore and Wiedenbeck, 1999) consider only a single type of task. However, considering recent studies of read-to-recall vs. read-to-do tasks in text comprehension (Mills et al., 1995; Schmalhofer & Glavanov, 1986), we may expect different levels of development of the program and situation models depending on the task, or purpose for reading a program (Détienne, 1996; Détienne, Rouet, Burkhardt and Deleuze-Dordron, 1996).

## 4. Research Questions

In this research, we examine the effect of expertise, task and phase on program comprehension.

Our first research question is how expertise in programming affects the construction of the two representations. In this study we do not manipulate expertise in the problem domain (all participants have knowledge in this domain), but rather we manipulate expertise in the programming domain. We compare expert programmers in OO programming with advanced computer science students learning OO programming. According to our model, the expertise of participants should not affect the construction of the program model, provided that our novices are advanced students, because both have text structure knowledge. However, expertise should affect the construction of the situation model because experts have much more knowledge of plans and the relationship of plans and objects.

A second research question is the effect of the task on the construction of mental representations in OO program comprehension. One hypothesis is that the read-to-recall purpose for reading focuses the understanding activity on the construction of the textbase, whereas the read-to-do purpose for reading focuses the understanding activity on the construction of the situation model. To study this question, we will analyse the evolution of these two types of representation depending on the task. We chose two tasks, a reuse task and a documentation task, because these are two realistic purposes for reading a program. Reading a program written by someone else for documenting it (as opposed to documenting while designing) can be considered similar to the read-to-recall task because it involves summarization more than problem solving (Détienne, 1996). This makes it most similar to text summarization (Schmalhofer and Glavanov, 1986) for text understanding. As in a text summarization task, programmers who read a program to document it are expected to concentrate on encoding the program text itself, i.e., constructing a textbase representation. On the other hand, in a read-to-do situation the programmer reads a program in order to use it for performing a problem solving task. In the reuse situation, the task is to design and implement a new program from the basis of the existing source program. We expect that, as in text understanding studies on read-to-do tasks, the reuse task entails the construction of a situation model of the source (Détienne, 1996). To summarize, our hypothesis is that the documentation task focuses the understanding activity on the construction of the program model, and the reuse task focuses the understanding activity on the construction of the situation model. As no studies in the mental models framework have investigated the interaction of task and expertise, this will also be investigated.

A third question is how the program model and the situation model are developed according to the stage of comprehension. Pennington found that the program model emerged earlier than the situation model in procedural program understanding. Our question is how the program model and the situation model each evolve as the programmer works with a program over time.

In order to investigate these research questions we conducted an experiment on OO program understanding by experts and novices, carrying out documentation or reuse tasks. Our experimental paradigm is similar to that used by Pennington. Participants answered questions after having studied a program in each of two time frames. The question categories were revised in accordance with our extension of Pennington's model.

## 5. Design and methodology

### 5.1 Experimental design

A four-factor mixed design was used, as shown in Figure 1. The between participants factors were expertise (OO expert vs. OO novice) and task orientation (documentation vs. reuse). The within participants factors were phase (initial study phase vs. task performance phase) and type of model tested (program model vs. situation model). Data came from correctness of responses to questions.

INSERT FIGURE 1 ABOUT HERE

### 5.2 Participants

Fifty-one participants took part in the experiment. Thirty were object-oriented experts and 21 were object-oriented novices. Thirty participants were recruited in the United States and were English language participants (16 experts/14 novices), while 21 were recruited in France and were French language participants (14 experts/7 novices). The experts were recruited by various means including electronic advertisements, contacts with software development enterprises, and nomination by colleagues. The novices were recruited through announcements in their universities.

The experts were professional programmers experienced in object-oriented design with C++. The mean age of the experts was 29.2 years and all but one were male. Their mean amount of programming experience (student and professional) was 9.8 years. The average amount of time for which they had used the C++ language was 3.3 years. The self-reported number of languages which they used or with which they were familiar was 8.5.

Our objective was to recruit novices who had sufficient programming experience to work with large programs, but had little experience in OO design. Therefore, the novices were chosen from advanced undergraduate computer science students who were experienced in C and other non-OO languages but had only basic knowledge of object-oriented programming and C++. All of these participants were currently or had been recently enrolled in a course introducing object-oriented programming and C++. The mean age of the novices was 26.8 years. Seventeen were male and 4 were female. The average amount of student plus professional programming experience was 5.15 years. The novice participants reported an average of .97 years of use of C++. The mean number of programming languages reported was 7.35.

## 5.3 Materials

The materials consisted of a database program of approximately 550 lines which managed personnel, student, and course information for a small university (See Burkhardt, 1997, for the full text of the program). The problem knowledge used in this program, e.g., management of personnel in an University, was supposed to be familiar to our participants, whatever their expertise in the programming domain. While small by industrial standards, this program was larger than programs typically used in similar research. More importantly, the program was large enough to include delocalized plans and program macrostructure. The program was a typical object-oriented C++ application. It was composed of 10 classes presented in 23 files, using a conventional arrangement in which each class had a .h file containing the class declaration and a .cc file containing the implementation of the functions of the class. Several files, such as main.cc, did not correspond to classes. The hierarchy of classes of this program is presented in Figure 2. The source program with which the participants worked took full advantage of OO features of classes, encapsulation, inheritance, composition of classes, function overloading, operator overloading, and polymorphism. As in Pennington's study (1987a), little documentation was included in the text of the program. The program existed both on-line and in hardcopy.

INSERT FIGURE 2 ABOUT HERE

During the task performance phase, reuse participants were given a variation of the library problem (Wing, 1988) to design and implement. The problem required the participant to create a database application to manage a small library (see the library problem statement in Appendix 1). Functional requirements included storing and retrieving information about books and journals, managing information about the different classes of users of the library, and maintaining up-to-date records about the library materials checked out to individual users. This problem was partially isomorphic to the database program and allowed for reuse by inheritance or by template copying and modification. The structure of a canonical solution for this library problem is displayed in Figure 3. Documentation participants were asked to comment the code for the use of another programmer who would later maintain it.

INSERT FIGURE 3 ABOUT HERE

Our methodological approach was similar to the one followed in Pennington's study (1987a; 1987b). Two sets of yes/no questions were developed, one to be used at the end of each phase. With this methodological approach, different types of questions are included to test whether the corresponding categories of information make up the representation constructed by participants as a result of a comprehension phase. The use of yes/no questions with time constraints to perform the test guarantees that replies are based directly on the available representation constructed from the comprehension phase, not on extra reasoning processes. Furthermore, yes/no questions prevent ambiguities in interpreting responses.

The questions fit conceptually into two classes targeting knowledge making up the program model and the situation model. Three information categories reflected information composing the program model (see Table 3). Six information categories reflected information composing the situation model (see Table 4). Given the size of our program, we controlled that, in each category, questions were related to distinctively distributed parts of the program. The difficulty of the questions was assessed in a pilot test. Based on the pilot test, we eliminated questions which were too difficult compared to others in a particular categories.

The two questionnaires were matched to each other. This was done by creating pairs of questions that were as similar as possible in the kind of information they asked for. For example, questionnaire 1 might ask "Does the program define a Professor class," while questionnaire 2 asked "Does the program define a Student class." Each questionnaire contained 54 questions (3 yes questions and 3 no questions in each of 9 question categories). Two versions, a French and an English versions, of each questionnaire were created. Within each questionnaire two forms, A and B, were created with different random orderings of the questions. An example of questionnaire is given in Appendix 2.

INSERT TABLE 3 & TABLE 4 ABOUT HERE

## 5.4 Procedure

Participants took part in a single experimental session lasting about 4 hours. The participants were run in the laboratories of one of the experimenters. Experts and novices were assigned randomly to the documentation or reuse groups. Each participant was run individually with an experimenter in the room.

The experiment was divided into two phases: phase 1 was the initial study of the program and phase 2 was the task performance phase. At the beginning of phase 1, participants were given a task orientation to study the program to later either reuse or document it, as appropriate. Participants were then given the database program and asked to study it for 35 minutes. They had access to hardcopy and also to an online version. During phase 1, participants were allowed to study the program, run it (an executable was provided), and take notes. They were not allowed to modify the program. A reference text for C++ was available (Stroustrup, 1991). Verbal protocols were collected. After this initial study phase, participants answered their first question set on-line during a maximum of 45 minutes. The order of the questionnaires was counterbalanced, so that approximately half the participants answered questionnaire 1 (either Form A or Form B) and half questionnaire 2 (either Form A or Form B) in phase 1. Participants were instructed to answer the questions as quickly as possible without making too many errors. They were given a maximum of 30 seconds to answer, after which they were automatically timed out. The questions were presented in 3 blocks of 18 questions each. The time between questions was 1.5 seconds. At the end of each block the participant had the option of resting. Questioning resumed when the participant pressed a button. The participants responses (Y, N, or T for timed out) and reaction times to the questions were recorded. At the end of phase 1 the participants took a break for approximately 10 minutes.

In phase 2, the task performance phase, participants were asked to carry out the documentation or reuse task for 90 minutes. They worked on the tasks they were initially assigned. Participants were given written task instructions, including the statement of the reuse problem for the reuse participants. The same materials were available as in phase 1: an on-line version of the program, the executable, a hardcopy program, and the textbook. Participants were allowed to take notes, run the program, modify the program, and create new files of their own. Documentation participants were instructed to place their documentation in the on-line files. Reuse participants were instructed to both design and implement their solution. The participants were told that they might not finish the task in 90 minutes but should make as much progress as possible. Again verbal protocols were collected. Finally, participants answered the second of their pair of comprehension questionnaires during a maximum of 45 minutes.

Instructions given to the participants are shown in Appendix 3.

At the end of the experiment participants filled in a demographic questionnaire and a questionnaire probing their approach to the documentation or reuse task. This questionnaire took about 15 additional minutes.

The timing of the experiment is summarized in Appendix 4.

# 6. Results

In this section we first present the results of several a posteriori tests used to evaluate the sources of variation in our experimental design. Then we present the main results with respect to the factors under study with a main emphasis on expertise, task, and phase. Our analysis involves the correctness of responses to the questions. We do not report results from the reaction times in this paper. Several results from the protocol analysis are presented to aid in understanding the behavior of participants during the documentation and reuse tasks.

## 6.1  A posteriori tests

The a posteriori tests included tests of the effect of the order of presentation of questionnaires, the order of questions within the questionnaires, the language of the questionnaires, and the number of non-responses. The results of these tests are summarized in the following paragraphs.

The order of presentation of the questionnaires had a significant effect on the total score of the participants, as indicated by the interaction of phase and questionnaire ($F(1,45)=23.258$, $p<.0001$). In analyzing phase 1, it was found that questionnaire 1 was responded to less correctly than questionnaire 2, regardless of expertise or task (mQuestionnaire1=33.778; mQuestionnaire2=37.5; $F(1,43)=9.291$, $p<.0039$). An analysis of phase 2 showed that there was no significant effect of the questionnaire. Because of the significant effect in phase 1, the order of presentation was used as a factor in the global analysis which follows. However, there it was found that it did not have a significant main effect and did not participate in any significant interactions, so it is not reported. The order of questions within the questionnaires (two different random orders) did not have a significant effect on the results. Therefore, this factor was not included in the following analyses.

The language of the experimental materials (English or French) did not have a significant effect on scores for questionnaire 1, nor for questionnaire 2. Nevertheless, analyzing each question separately, we found two questions on which the responses were systematically different for the English and French versions. These two questions were eliminated from the analyses. Other sources of variation were the two experimenters and the two laboratories where the experiment was run. We can only measure their effect indirectly through the language factor. However we ran the experiment according to a standardized script.

Our analyses showed that there were very few cases in which the participant failed to respond within the 30 seconds allowed (42/5508=.0076 percent). A substantial number of participants

(17) timed out on at least one question. Although most of these participants had 1-5 time outs, one participant was responsible for 15 (41%) of the time outs. The data from this participant were eliminated from further analysis. For the remaining non-responses we assigned a yes response randomly to half and a no response randomly to half.

## 6.2 Main effects and interactions

A four-way mixed model Analysis of Variance was performed on the number of correct responses to questions. The between subjects factors were expertise (novice or expert) and task (documentation or reuse). The within subjects factors were phase (1=initial phase or 2=task performance) and type of model (program model or situation model). We based all of our analyses on the percentage of correct responses to questions.

There was a significant overall effect of expertise (mexpert=67.824, sd=12.875; mnovice=63.858, sd=10.990; $F(1, 46)$=6.836, p<.0120). As expected experts are better than novices at constructing correct mental representations of the program. There was a significant overall effect of task (mreuse=64.685, sd=12.419; mdoc=67.920, sd=11.971; $F(1, 46)$=4.230, p<.0454). Overall, the correctness of the mental representation was higher in the documentation condition than in the reuse condition. The results showed that there was a significant effect of phase (mph1=63.931, sd=12.164; mph2=68.544, sd=12.023; $F(1, 46)$=15.469, p<.0003). Overall, the correctness of the mental representation improved between phase 1 and phase 2. There was a significant overall effect of type of model (msituation=73.086, sd=9.597, mprogram=59.389, sd=10.789; $F(1,46)$=99.333, p < .0001). Overall, the subjects responded more accurately to the questions related to the situation model than those related to the program model. However, we cannot interpret directly these results in terms of dominance of one model overt the other because the difficulty of the questionnaire cannot be calibrated.

The two-way interaction of expertise and type of model was significant ($F(1,46)$=5.659; p<.0216) ). The situation models of experts and novices differed (smexpert=75.926, sd=8.936, smnovice=68.826, sd=9.054) but their program models did not (pmexpert=59.722, sd=10.999, pmnovice=58.889, sd=10.586). The other two-way interactions and the three-way interactions were not significant. However the four-way interaction of expertise, phase, task and type of model was significant ($F(1,46)$=5.658, p<.0216). The results are shown graphically in Figure 4. See Table 5 for the means of the interactions.

INSERT TABLE 5 AND FIGURE 4 ABOUT HERE

The existence of the four-way interaction encouraged us to decompose our data according to the task factor. This allows us to examine more deeply how the mental representations evolved in documentation and in reuse.

For the documentation group, the effect of expertise was significant (mexpert=69.926, sd=12.742, mnovice=65.111, sd=10.307; $F(1,22)$=4.427, p<.0470 ). The effect of phase was significant (mph1=65.127, sd=11.639, mph2=70.712, sd=11.758; $F(1,22)$=10.405, p<.0039). The effect of type of model was significant (msituation=73.918, sd=9.211, mprogram=61.921, sd=11.461; $F(1,22)$=45.520, p<.0001). We found a two-way interaction between type of model and expertise ($F(1,22)$=6.873, p <.0156). Experts and novices differed in their situation models (smexpert=77.748, sd=8.698, smnovice=68.556, sd=7.103) but not in their program models (pmexpert=62.103, sd=11.320, pmnovice=61.667, sd=11.943). This confirms, for the documentation group, our hypothesis that the expertise of programmers should affect the construction of the situation model but not the construction of the program model, provided that our novices are advanced students. There was no three-way interaction.

For the reuse group, there was no effect of expertise. We found a significant effect of phase (mph1=62.826, sd=12.640, mph2=66.544, sd=12.029; F(1,24)=5.415, p<.0287) and of type of model (msituation=72.318, sd=9.967, mprogram=57.051, sd=9.661; F(1,24)=55.381, p<.0001). However, there was a three-way interaction between phase, expertise and type of model (F(1,24)=6.110, p<.0209).

Given the three-way interaction, we decomposed the reuse group further into novices versus experts in the reuse situation. There was no effect of phase for either novices or for experts. If we consider experts and novices separately, we find an effect of type of model for both groups (smnovice=69.097, sd=10.847, pmnovice=56.111, sd=8.435; F(1,9)=48.856, p<.0001) (smexpert=74.332, sd=8.969, pmexpert=57.639, sd=10.441; F(1,15)= 32.008, p<.0001).

However, for the novice group we found a two-way interaction between model and phase (F(1,9 7.495, p<.0229). In this group, the effect of phase was to increase the construction of the situation model but not of the program model. For the novices in the reuse situation there was a significant increase of the situation model between phase 1 and phase 2 (mnovice/reuse/ph1=63.028 sd=10.382; mnovice/reuse/ph2=75.167 sd=7.664; t(9)=-2.657, p<.0262). For the expert group, there was no such two-way interaction.

## 6.3 Effect of expertise on documentation task performance

For the documentation group, we have found support, based on the participants' answers to questions, for the hypothesis that the expertise of programmers should affect the construction of the situation model but not the construction of the program model. A question then is how this differential construction of representations is reflected in the documentation task and, more particularly, in the kind of comments produced by these two groups of programmers. In order to address this question, we have analysed the comments made by participants, as a result of phase 2 (task realization), in function of their expertise. Our working hypothesis is then that comments reflect the kinds of representation constructed by participants from the program text and from their own knowledge.

Globally, the number of comments produced by experts and novices is not significantly different. We have analyzed the nature of the parts of program commented. We distinguished comments located in .h files, .cc files and outside of files. The class header file (.h file) is more abstract than the implementation file (.cc file) inasmuch as the header file contains the class declaration including declarations of attributes and methods, while the code implementation of methods is made in the .cc file. There was a significant overall effect of location of comments (hmean=22.045, sd=26.072, ccmean=18.227, sd=20.426, outsidemean=2.955, sd=6.890; F(2,36)= 5.857, p<.00063). We found a two-way interaction between expertise and location (F(2,36)=3.260, p<.05). As shown in Figure 5 (see details in Table 6), expertise affects the location of comments: experts make more comments related to .h files whereas novices make more comments related to .cc files. In order to refine this analysis, we have distinguished comments by the nature of objects commented : class, function, inline, file (as shown in Table 7). In .h files, experts as well as novices produce more comments in relation to functions or goals. In .cc files, most comments produced by experts are related to functions whereas most comments produced by novices are related to inline code.

Given that .h files contain more abstract information about the program than .cc files, and that comments related to functions or goals reflect the situation model, these results may reflect the different construction of the situation model in function of expertise.

INSERT FIGURE 5 ABOUT HERE

INSERT TABLE 6 and TABLE 7 ABOUT HERE

While these results are consistent with the kinds of representation constructed as a function of expertise presented in the previous section, they are different from those found in the literature on documentation (Riecken et al., 1991; Rouet et al., 1995). We argue that the documentation task results of these prior studies, interpreted as a read-to-recall task, show that the documentation task biases experts toward the construction of a program model, as reflected in the comments produced. A major difference between our study and these previous studies is the size of the program. The large size of our program may have prevented the participants from constructing a more complete program model.

## 6.4 Effect of expertise on reuse task performance

The analysis of the correctness of answers on questions shows that for the reuse participants, there was no global effect of expertise, but there were global effects of phase and type of model. Both novices and experts responded more accurately to the question set on the situation model than to the question set on the program model, but we cannot directly interpret this result as superiority of one model over the other given our methodology. An interesting result is that the situation model improved during phase 2 only for the novice group and became comparable to the situation model of the experts. To better understand these results, the characteristic of the activities carried out by the novice and expert groups during the reuse has to be considered.

Based on the protocols recorded during the task performance phase, the analysis of the subjects' performance in the reuse group shows that they actually spent a significant amount of time in reuse activities. During this phase, which lasted 90 minutes, the novices spent a mean of 49 minutes carrying out reuse activities ( sd= 21 minutes), and the experts spent a mean of 52 minutes (sd= 12 minutes). There was no significant difference between novices and experts ($t(23) = .521$, p<.6074). As is often the case in novice-expert comparisons, the variability within the novice group was greater.

We investigated the prevalence of reuse by cut and paste and also reuse by inheritance. Our motivation for this investigation is that some prior work on reuse in the OO paradigm (Lange and Moher, 1989) has found reuse almost entirely by cut and paste, whereas reuse by inheritance would be expected because it is supported by the OO paradigm. We expected that reuse by cut and paste would be widely used. However, we also expected to find a substantial amount of reuse by inheritance in a well structured, hierarchically organized OO program. These expectations were confirmed. Reuse by cut and paste was globally most prevalent, used in 46 percent of the reuse episodes. The percent of use of cut and paste during the reuse episodes did not differ statistically between experts and novices (mexpert=52.20%, sd=23.4; mnovice= 37%, sd=21; $t(23) = 1.656$, p<.1112). Reuse by inheritance was also widely used by both novices and experts in 25.71 percent of the reuse episodes. There was no significant difference between experts and novices in episodes of reuse by inheritance (mexpert=20.90%, sd=21.90%; mnovice=33%, sd=23.70%; $t(23)=-1.309$, p<.2033).

We also analyzed how many classes were reused and which classes tended to be reused by experts and novices. For the analysis of number of classes reused, we divided the results into classes from the source program reused in the target program (inter-program reuse) and classes created in the target program and then later reused within the target program (intra-program reuse). The mean for reuse between the source and target program was 6.72 classes, while the mean for reuse within the target program was 5.88. There was no significant difference between novices and experts in either type of reuse. However, there were differences in which classes were reused heavily by experts and novices. Taking classes reused by 70 percent of participants as a cut-off, we found that in inter-program reuse experts reused most heavily two key classes from the source program. The class Collection, where most of the database manipulations were implemented, was reused by 100 percent of expert participants, while the class Example, which served as the generic top of the class hierarchy, was reused by 87 percent of the experts. The classes Employee and String were both reused by 73 percent of the experts. Seventy percent of novices reused the key classes, Collection and Example. The most frequently reused class among the novices was the class String, reused by 90 percent, which contained stereotypical string manipulation functions. In the reuse of classes within the target program, experts reused classes that were related to each other in the inheritance hierarchy, for example the class Document and its children, Book and Periodical. Within the target program, novices reused heavily only the classes Book and Periodical. Unlike the experts, their reuse did not correspond to the hierarchical inheritance structure of the program.

These results suggest that both experts and novices are able to select the most pertinent classes for reuse between the source and target program. However, it appears that novices may be somewhat less adept than experts at identifying classes for reuse. This would be consistent with the results of Lee and Pennington (1994) that OO novices are less adept at identifying objects in an OO design task. However, we point out that the differences in our study were small, and also that identifying objects from a problem statement is not the same task as identifying the usefulness of an existing class for reuse. In intra-program reuse there appeared to be a clear difference between experts and novices in their reuse of hierarchical inheritance structures.

## 7. Discussion

In this section, we discuss the role of expertise and task in the development of the mental representations, as well as their interactions. We also discuss changes in the mental representations over the two phases. Finally, the implications of our results are developed.

## 7.1 The effect of expertise in the mental model approach to OO program understanding

Previous studies in the mental model approach to text understanding have shown a differential effect of expertise on the construction of the situation model and program model. Similarly, according to our mental model approach to OO program comprehension, the expertise of programmers should affect the construction of the situation model but not the construction of the program model. This result was found for the documentation group as shown by the two-way interaction between type of model and expertise. For this group, experts and novices differed in their construction of the situation model but not of the program model. By contrast, in the reuse situation, we did not find such an interaction.

Thus, we replicate the results found on the effect of domain knowledge in the text comprehension literature only in the documentation situation. This leads us to consider, a posteriori, this situation as a control situation with respect to the task factor because the task of documentation does not affect the representations habitually constructed by readers as a function of their expertise.

## 7.2 The effect of task in the mental model approach to OO program understanding

Previous studies, in the mental model approach to text understanding, have shown a differential effect of task, or purpose for reading, on the construction of the situation model and the program model. Authors distinguish between the read-to-recall purpose for reading, which focuses the understanding activity on the construction of the textbase, and the read-to-do purpose for reading, which focuses the understanding activity on the construction of the mental model. Similarly, we expected the documentation task to focus the understanding activity on the construction of the program model and the reuse task to focus the understanding activity on the construction of the situation model. Our results showed a global effect of the task factor and, more interestingly, a complex four-way interaction of expertise, phase, task and type of model.

In the documentation task, we found effects of expertise, phase and type of model. We also found a significant interaction between expertise and type of model. The expertise of programmers affects the construction of the situation model but not the construction of the program model. We expected the documentation task to be similar to a read-to-recall purpose for reading. As in a text summarization task, programmers who read a program to document it were expected to concentrate on encoding the program text itself, i.e., constructing a textbase representation. While we found a significant effect of phase and of type of model, we did not find an interaction between type of model and phase, which would have revealed a differential construction of the program model and the situation model over time. In fact, both models evolved over time, which is probably due to the extra time (phase 2) spent reading the program in this task. A posteriori, we consider this task as a control situation which is neither read-to-recall-like nor read-to-do-like. Rather, the results show the classical effect of domain knowledge on the construction of mental representations. Our results are different from those found in the literature on documentation (Riecken et al., 1991; Rouet et al., 1995). The difference may be explained by the much larger size of our program, which may have prevented the participants from constructing a more complete program model.

In the reuse task, we expected that, as in text understanding studies on the read-to-do purpose for reading, the reuse task would focus the understanding activity on the construction of the situation model. We found a three-way interaction between phase, expertise and type of model. Furthermore, for the novice group we found a two-way interaction between type of model and phase. In this group, there was an increase of the construction of the situation model over time but not of the program model. This is what was expected for a read-to-do-like task. However, it was observed only in the novices, not in the experts.

Thus, an interesting result is that the reuse task appears to entail a decrease of the expert/novice differences as concerns the construction of the situation model. This suggests that, even though novices have been shown to fail to develop an accurate situation model (as found for novices compared to experts in the documentation-control situation), they are capable of building a situation model if they are given a task that requires situation knowledge.

For the expert group, an interpretation of the results may be that their situation model constructed in phase 1 was already developed enough to perform the reuse task. Also, the reuse task is a double-task situation in which experts could choose, in phase 2, to spend more time solving the target problem rather than simply continuing to understand the source. This would be consistent with "as-needed" (Littman et al., 1986) or "relevance" (Koenemann and Robertson, 1991) strategies of comprehension.

Our results seem, at first sight, contradictory to the idea, developed by Lange and Moher (1989) that programmers tend to use a strategy of "comprehension avoidance" of the source in a reuse task. These authors found comprehension avoidance in an experiment in which a single programmer developed a large program in Objective-C by extensive use of copy and paste. We argue that the situation observed by Lange and Moher is quite different from our experimental situation on two dimensions: the intra- versus inter-program reuse situation and the familiarity of the programmer with the source. In Lange and Moher's observational study, reuse was intra-program and the programmer was familiar with this source as he had developed it himself. In this case, comprehension avoidance may only mean that the representation already constructed from the source was sufficient to reuse some parts of it without having to read the code in detail again. In our experiment, the source had obviously been produced by somebody else and this was mostly an inter-program reuse situation (even though intra-reuse inside the target program was allowed). In this situation, comprehension of the source is probably a step necessary to the reuse of it.

Our experimental situation is in fact more similar to the Rosson and Carroll's study (1993) on reuse of classes in Smalltalk. In that study, programmers were encouraged to reuse existing classes, with which they were not familiar, in order to modify a program. These authors observed that the programmers made extensive use of an example program, using these classes. Rosson and Carroll interpret this as a way to understand the source classes. Similarly in our experimental situation, comprehension of the source seems a step necessary to the reuse of it.

To go further in our interpretation of the effect of the reuse task, we can also discuss it in the framework of analogical reasoning. Gick and Holyoak (1983) distinguish two conceptually distinct ways in which a schema could be involved in solving a problem with reference to information obtained from prior analogs. In the case referred to as "reasoning from an analog", the new problem is mapped directly to a prior analog to generate an analogous solution. A schema does not exist as an individual concept prior to this mapping. However, schema induction may be the result of analogous reasoning. This situation is close to those studied in case-based reasoning. In the case referred to as "reasoning from a schema", a schema has already been induced from prior analogs and stored in memory. Therefore, the participant can directly reason from the schema and instantiate the schema so as to elaborate a new analog.

The reuse situation of novices could be classified in the first category. In this case, analogical reasoning would entail the construction of conceptual knowledge that would account for both the source and the target. This construction would be reflected in the representation of the source, in particular in the evolution of the situation model. The reuse situation of the experts, on the other hand, could be classified in the second category.

## 7.3 Changes in the mental representations in the two phases

For procedural program comprehension, Pennington found that the program model emerges earlier than the situation model in program understanding. In this study we did not directly compare the two models. However, we can discuss the evolution of each model separately over the phases. There was a global effect of phase. However, the analyses showed that there was a significant evolution of the situation model only for novices in the reuse condition. Thus, the evolution of the type of model seems related to both expertise and the task.

## 7.4  Implications of this research

In the reuse task the difference between experts and novices on the situation model decreased and even disappeared over the course of the experiment. The reuse task seemed to focus the understanding activity of novices on the construction of the situation model. Ultimately, their situation models reached the same level of development as those of the experts in this task. This result has important instructional implications. It suggests that, motivated by some tasks, low domain knowledge readers may construct a representation that they do not spontaneously construct otherwise. More specifically, if we give low knowledge domain readers a read-to-do purpose for reading, they are able to construct a situation model that is comparable to the one construct by high domain knowledge readers.

## 8.  Limitations and future directions

There are several limitations of this study related to the methodology. Most importantly, we cannot validly compare the situation model and the program model because it is difficult to calibrate the difficulty of the questions making up the two models. We believe that developing a calibration of the difficulty of the questions is a prerequisite to a comparison of models. While Pennington made such a comparison, she was working with very simple and small programs. Applying this method to a complex program makes the questions less comparable to each other. While it may be possible to develop comparison criteria for program model questions, it seems quite difficult to do this for situation model questions which are based on more extensive domain knowledge. This suggests that when a program reaches a certain complexity strict experimental methods reach their limit of applicability.

Several other limitations can also be identified. First, it should be noted that phase was confounded with task orientation vs. task performance. Thus, it is not possible to determine whether the changes observed in phase 2 were the result of performance of the task or of additional time to study the program. Second, participants worked with a single program which implemented a database. To generalize the results it is necessary to repeat the study with other programs in other problem domains. Third, while the program was larger than often used in this kind of study, it was still a small program by industrial standards. Thus, we do not know whether the mental representation of a much larger program would conform precisely to what we found here. Fourth, in our study participants worked with the program for approximately 2 hours, and most did not have time to finish the reuse or documentation task they were given. We might have observed further evolution of the mental representation if they had worked with the program over a longer time. Finally, this study should be repeated with other object-oriented languages to eliminate the possibility that the procedural components of C++ unduly affected these results.

A direction of research would be to distinguish expertise in the task from expertise in domain knowledge. Two studies (Rouet et al., 1995; Woodfield, Embley & Scott, 1987) have shown that experts in domain knowledge are not necessarily experts in the reuse task. Based on these results, we could expect the effect of the task to be more complex that what we found.

Two theoretical questions are raised by this research. First, a two-level model may be sufficient to account for comprehension of small programs, such as those used in earlier studies. However, we believe that such a model may be too simplistic to describe the situation model of complex programs. An extended model should probably distinguish between multiple levels of abstraction, including levels that are more inference-based and independent of program implementation. Developing an extended situation model will be difficult because it will depend on the domain of the program and the task. We believe that important theoretical advances in this domain will require two steps. The first step will be to develop more empirical studies of large program comprehension and software system comprehension. At that point we move from the study of programming-in-the-small to programming-in-the-large. A second step will be to integrate theoretical advances made in these two branches of the field.

The second theoretical question concerns the generality of the proposed comprehension model. The program model is highly dependent on the notation of the specific language. We expect that information that is highlighted in the notation will be easier to extract and will be more fully represented in the program model. Future revision of the program model components of our proposed model should take into account the strong links of these ideas with studies of notational structure (e.g., Gilmore and Green, 1984; Green et al., 1991). While we expect similar kinds of information to be represented in the situation model regardless of the language, in particular static and dynamic information referring to objects and plan knowledge (Rist, 1995), the representation may be easier or harder to construct depending on the notation, and thus on the program model.

| Text relations | Knowledge structures | Mental representation | Model |
|---|---|---|---|
| Elementary operations | Text structure knowledge | Dynamic and functional views | Program model |
| Control flow | Text structure knowledge | Dynamic view | Program model |
| Main goals | Plan knowledge | Functional view | Situation model |
| Data flow | Plan knowledge | Dynamic and functional views | Situation model |

Table 1 Correspondence between text relations, knowledge structures, mental representation, and model in procedural program understanding (adapted from Pennington, 1987b)

| Text relations | Knowledge structures | Mental representation | Model |
|---|---|---|---|
| Control flow | Text structure knowledge | Dynamic view | Program model |
| Elementary operations | Text structure knowledge | Dynamic and functional views | Program model |
| Elementary functions | Text structure knowledge | Dynamic and functional views | Program model |
| Program objects | Problem knowledge and plan knowledge | Object view | Situation model |
| Relations between program objects | Problem knowledge and plan knowledge | Object view | Situation model |
| Reified objects | Generic programming knowledge and plan knowledge | Object view | Situation model |
| Main goals | Problem knowledge and plan knowledge | Functional view | Situation model |
| Client-server | Plan knowledge (complex delocalised plans) | Dynamic and functional views | Situation model |
| Data flow | Plan knowledge | Dynamic and functional views | Situation model |

Table 2 Correspondence between text relations, knowledge structures, mental representation, and model in OO program understanding

| **Elementary operations** | Does the program contain the code fragment:<br>if ((number == search) ‖ (name == search)) return TRUE; else return FALSE;? |
|---|---|
| **Control flow** | In "initialize" are the professors initialized before the courses ? |
| **Elementary functions** | Does "Collection::maintain" print out a list and ask the user to input a selection.? |

Table 3 Example of questions from categories composing the program model

| Problem objects | Does the program define a "Schedule" class? |
|---|---|
| Computing objects | Does the program define a "Collection" class? |
| Object relationships | Does the "Researcher" class inherit from the "Employee" class? |
| Goals | Does the program allow you to create a new schedule for an upcoming semester? |
| Client-server | Does the "Schedule" class call a member function of the "Course" class? |
| Data flow | In "Schedule::maintain" does the value of "selection" affect the value of "offerings"? |

Table 4 Example of questions from categories composing the situation model

| | Documentation | | | | Reuse | | | |
|---|---|---|---|---|---|---|---|---|
| | **Expert** | | **Novice** | | **Expert** | | **Novice** | |
| | Program Model | Situation Model | Program Model | Situation Model | Program Model | Situation Model | Program Model | Situation Model |
| **Phase 1** | 59.921 (10.491) | 73.710 (6.981) | 57.778 (14.628) | 67.750 (6.300) | 55.556 (10.924) | 73.125 (11.248) | 57.778 (8.765) | 63.028 (10.382) |
| **Phase 2** | 64.286 (12.076) | 81.786 (8.561) | 65.556 (7.314) | 69.361 (8.086) | 59.722 (9.834) | 75.538 (6.051) | 54.444 (8.198) | 75.167 (7.664) |

Table 5 : Means (and standard deviations) of the interactions (percentage of correctness)

|  | Expert | | Novices | |
|---|---|---|---|---|
| **.h files** | 29.92 | (30.286) | 10.67 | (12.728) |
| **.cc files** | 9.92 | (9.535) | 30.22 | (26.171) |
| **outside** | 4.92 | (8.529) | 0.11 | (0.333) |

Table 6 : Means (and standard deviations) of number of comments produced by the experts and novices in function of their location during the documentation realization phase.

| | Experts | | | Novices | | |
|---|---|---|---|---|---|---|
| | .h files | .cc files | outside | .h files | .cc files | Outside |
| **Class** | 17,48% | 3,88% | 84,38% | 12,50% | 0,00% | 100,00% |
| **Function** | 49,61% | 51,16% | 14,06% | 56,25% | 32,35% | 0,00% |
| **Inline** | 29,82% | 29,46% | 0,00% | 30,21% | 66,91% | 0,00% |
| **File** | 3,08% | 15,50% | 1,56% | 1,04% | 0,74% | 0,00% |

Table 7 : Percentage of comments produced by experts and novices for each type of files, in function of the type of comments during the documentation realization phase.

|  | Documentation | | Reuse | |
|---|---|---|---|---|
|  | **Expert N=14** | **Novice N=11** | **Expert N=16** | **Novice N=10** |
| **Phase 1** **Study** | Study pgm for documentation, then comprehension questions | Study pgm for documentation, then comprehension questions | Study pgm for reuse, then comprehension questions | Study pgm for reuse, then comprehension questions |
| **Phase 2** **Task** | Documentation task, then comprehension questions | Documentation task, then comprehension questions | Reuse task, then comprehension questions | Reuse task, then comprehension questions |

Figure 1 Experimental design

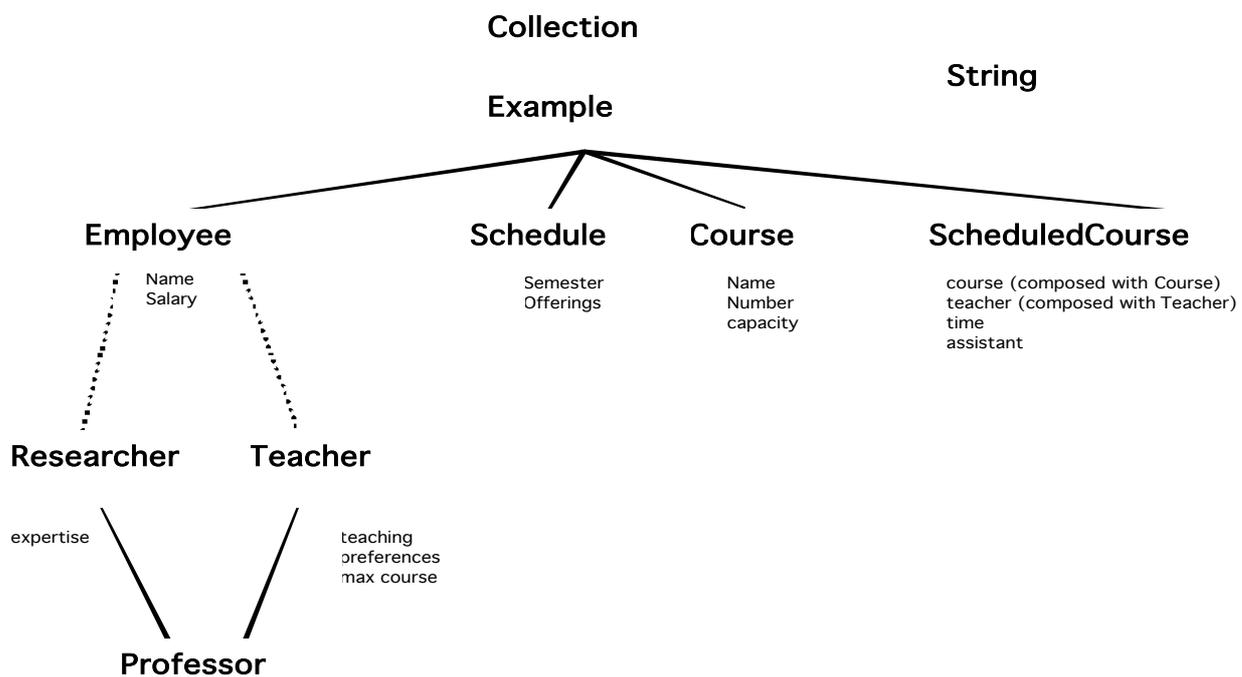

Figure 2 : Hierarchy of the database program

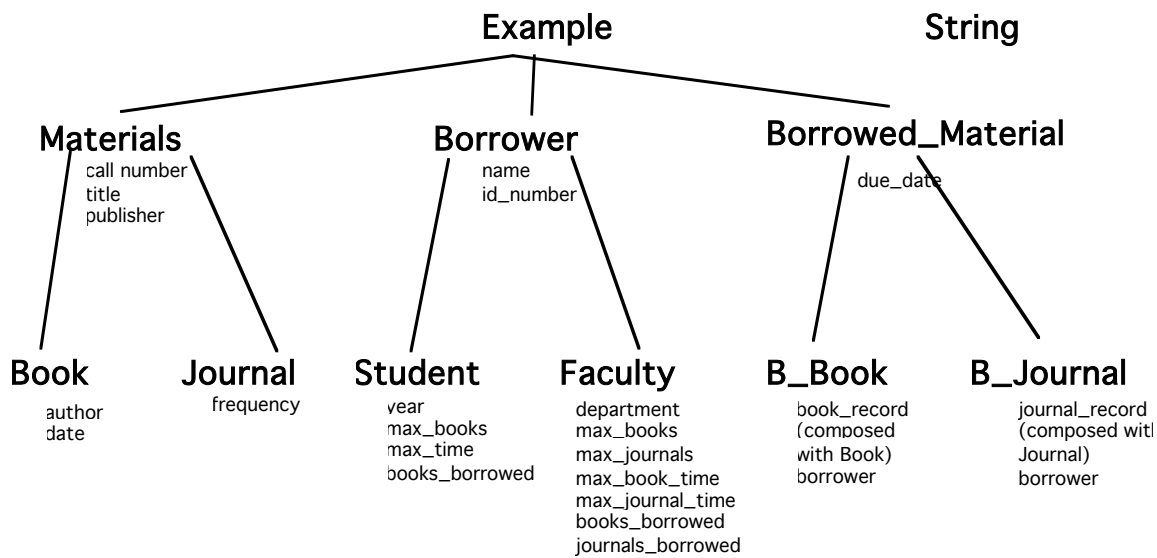

Figure 3 : Canonical solution for the Library problem

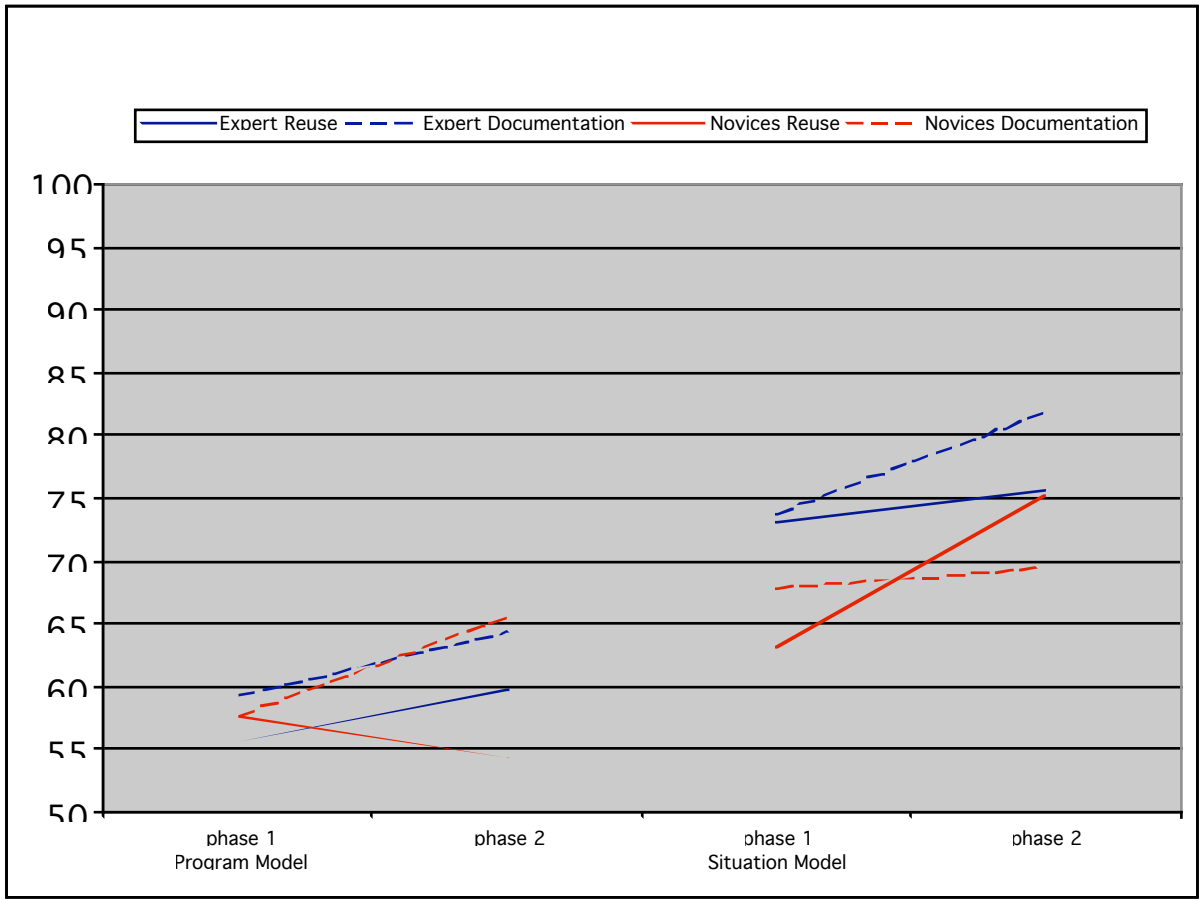

Figure 4. Effect of phase, expertise, and task on the level of
development of the program model and the situation model

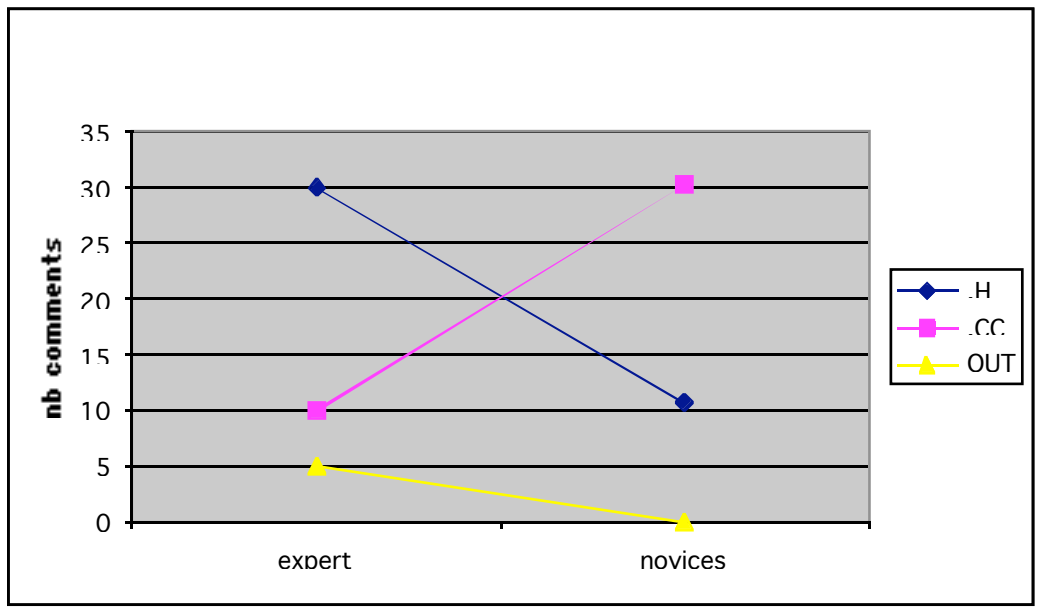

Figure 5. Effect of expertise on the number of comments produced as a function of location by the participants during the documentation realization phase

# APPENDIX 1

### The library problem statement

Write a program to solve the problem given below. You may write your own code as well as use any portions of the code of the program that you previously studied. Go about solving the problem as you normally would, but speak aloud as you work. The experimenter will prompt you if you forget to speak aloud.

This program will manage data about a small library. The library maintains three kinds of data: 1) data describing books and journals owned by the library, 2) data about the clients of the library and what books they have borrowed, and 3) data about books and journals that have been borrowed.

## Books and Journals

The library has books and journals. The bibliographic record of each **book** consists of the following:

> call number
> title
> author(s)
> publisher
> date of publication

You should be able to search for a book by author or title and add a record for it if it is not found. If it is found, you should be able to delete it or change its title or date.

The bibliographic record for each **journal** consists of the following:

> call number
> title
> volume
> number
> publisher
> frequency (e.g., monthly, quarterly...)

For journals you should be able to search by title and add a journal that is not found. If it is found, you should be able to delete it or to change its title or frequency.

## Clients of the Library

The clients of the library who may borrow materials are students and professors.
**Student** data consist of the following:

> name
> identification number (5 digits)
> year in school (1-4)
> maximum number of books allowed to be checked out SIMULTANEOUSLY:
> > (10 for student in years 1 and 2; 20 for students in years 3 and 4)
> maximum length of book loan (3 weeks)
> number of books currently checked out
> call numbers of books currently checked out

Note that students may **not** check out journals.

Data about **professors** consist of the following:

> name
> identification number (5 digits)
> department

maximum number of books allowed to be checked out
SIMULTANEOUSLY(50)
maximum number of journals allowed to be checked out
SIMULTANEOUSLY(10)
maximum length of book loan (26 weeks)
maximum length of journal loan (1 week)
number of books currently checked out
number of journals currently checked out
call numbers of books currently checked out
call numbers of journals currently checked out

You should be able to search for a client by name or identification number and add the client if not found. You should also be able to delete a client and to change a client's books checked out and journals checked out as books and journals are borrowed and returned. You should not allow clients to check out more books or journals if they have reached their maximum number. Also, do not allow students to check out journals.

**Borrowed Materials**

The library also keeps a record of all books and journals currently checked out.

For **books** this consists of:
the full bibliographic record of the book (as described above)
the name of the borrower
the identification number of the borrower
the due date

For **journals** this consists of:
the full bibliographic record of the journal (as described above)
the name of the borrower
the identification number of the borrower
the due date

You should be able to add books and journals to these records of checked out materials as they are borrowed and delete them when they are returned. You should also be able to generate a list of clients who have overdue books.

APPENDIX 2

**Example Questionnaire**

This example shows one questionnaire organized by model and type. Participants saw the questionnaire randomized and without labels for model and type

**Program Model**

*Elementary operations*

1. Does the program contain the code fragment: if ((sel_example = search (selection)) != NULL) ?
2. Does the program contain the code fragment: Example *element = list[0] ->alike(name); ?
3. Does the program contain the code fragment: if ((number == search) || (name == search)) return TRUE; else return FALSE; ?
4. Does the program contain the code fragment: bool Scheduled_course::check () { return (course->check)); ?
5. Does the program contain the code fragment: Example * select() {return TRUE} ?
6. Does the program contain the code fragment: if (preferences->search(c) != TRUE) ?

*Control flow*

7. In "initialize" are the professors initialized before the courses ?
8. In "Schedule::assign" is the time for a course selected before the teacher?
9. In "Teacher ::add" are a teacher's preferences and maximum number of courses checked before assigning a course to the teacher?
10. In "Collection::add_alike" is an element added to a list before the list is checked to see if it is full?
11. In "Course::alike" does the program ask the user whether he wants to add a course before searching to see if it exists?
12. In "Professor::all->maintain_T" are a professor's preferences printed out before the program asks whether the user wants to maintain the professor?

*Elementary functions*

13. Does "Collection::maintain" print out a list and ask the user to input a selection?
14. Does "Course::alike" return a course object that matches the string "name"?
15. Does "Schedule::assign" ask the user to input a name and search for a course that matches the name?
16. Does "Collection::remove" return a pointer to an empty  list?
17. Does "Teacher::add" add a new teacher?
18. Does "Scheduled_course::alike" search for a scheduled course and return either TRUE or FALSE?

**Situation Model**

*Problem objects*

19. Does the program define a "Schedule" class?
20. Does the program define an "Employee" class?
21. Does the program define a "Researcher" class?
22. Does the program define a "Student" class?
23. Does the program define a "Report" class?
24. Does the program define a "Time" class?

*Relationships between problem objects (inheritance and composition)*

25  Does the "Professor" class inherit from the "Teacher" class?
26. Does the "Researcher" class inherit from the "Example" class?
27. Is the field "Schedule::sched_course" of the type "Scheduled_course"?
28. Does the "Scheduled_course" class inherit from the "Course" class?
29. Does the "Teacher" class inherit from the "Researcher" class?
30. Is the field "Scheduled_course::teacher" of the type "Teacher"?

*Computing or reified objects*

31. Does the program define a "Collection" class?
32. Is the field "Course::number" of the type "String"?
33. Is the field "Teacher::preferences" of the type "Collection"?
34. Does the program define a "Set" class?
35. Does the "Employee" class inherit from the "Collection" class?
36. Is the field "Collection::list" of the type "String"?

*Main goals*

37. Does the program allow you to create a new schedule for an upcoming semester?
38. Does the program allow you to change the name of a course?
39. Does the program allow you to assign a teacher to a course which is being scheduled?
40. Does the program allow you to assign a teaching assistant to a course.
41. Does the program allow you to modify the maximum number of courses a professor may teach?
42. Does the program allow you to retrieve the number of students who are already registered in a course?

*Client-server relationships*

43. Does the "Professor" class call a member function of the "Teacher" class?
44. Does the "Schedule" class call a member function of the "Professor" class?
45. Does the "Professor" class call a member function of the "Course" class ?
46. Does the "Researcher" class call a member function of the "Example" class?
47. Does the "Teacher" class call a member function of the "Scheduled_course"class ?
48. Does the "Course" class call a member function of the "Employee" class ?

*Data flow relationships*

49. In "Collection::add_alike" does the value of "filled" affect the value of "element"?
50. In "Schedule::maintain" does the value of "selection" affect the value of "offerings"?
51. In "Course::maintain" does the value of "selection" affect the value of "capacity"?
54. In "Collection::add" does the value of "grow_increment" affect the value of "filled"?
53. In "Employee::identify" does the value of "name" affect the value of "search"?
54. In "Teacher::add" does the value of "c" affect the value of "preferences"?

# APPENDIX 3

## Instructions for Experiment

We need to give general instructions at the beginning of the experiment and before each part of the experiment. I think that the general instructions should be given verbally following a written script. The instructions for each part of the experiment should be given in written form on paper, with a separate sheet of paper for each part of the experiment.

I would like to give the general instructions verbally because we need to give some kind of welcome and overview. It would seem too impersonal to have no verbal interaction with a participant at the start of the experiment. In the initial instructions we also need to give instructions about the think-aloud method. Should these be given verbally or in written form (see below)?

The rest of the actual task instructions can be given in written form to ensure that everyone is given identical information. I think that it is better to give these instructions on paper rather than present them on the computer. Normally we will be using 2 computers--the participant's computer for doing the reuse/documentation task and our computer for answering the comprehension questions. It seems simpler to me to have the instructions on paper than to have to worry about presenting instructions on two different computers, uploading instruction files, etc.

### Introduction and Overview of the Experiment

Many claims have been made about the superiority of object-oriented programming. In this experiment we are attempting to evaluate those claims. We are particularly interested in the ease of comprehension of object-oriented programs and in their use and documentation.

This experiment will take approximately 3.5 hours to complete, and you will be allowed to take short breaks during that time, if you wish. The experiment is divided into 5 parts:

1. Study and comprehension of an existing object-oriented program
2. Comprehension questions about the program
3. Reuse of the existing program in the solution of another problem
       OR
   Documentation of the object-oriented program that you previously studied
4. Comprehension questions about the program
5. A questionnaire about your progamming background and the approach you used to carry out the task you were given in this experiment

You have been asked to participate in this experiment because you have experience in object-oriented programming. We wish to emphasize that our objective is to evaluate claims about object-oriented programming, not in any way to evaluate your skills.

Your participation will lead to a better understanding of the type and magnitude of benefits that can be expected from object-oriented programming.

Do you have any questions?

### Think-aloud Instructions

During the study of the program and the reuse or documentation phase we will be using a **think-aloud** method of data collection.

You will be asked to **"think out loud"** as you work. You should speak aloud all of your thoughts as you study the program and later do the reuse (documentation) task. We want you to say <u>everything</u> that passes through your mind as you work.

If you are looking for something, say that.

If you are confused about something, say that.

If you have a hypothesis about something, say that.

Also, each time you switch your attention from one file to another or within a file from one part to another (e.g., from one method to another) , please tell the experimenter. **You cannot say too much**, and everything that you think about is of interest, even if it is seemingly irrelevant.

If you are silent for more than 20 seconds or if you change activities, the experimenter will remind you to continue to speak aloud by saying "What are you thinking?" Your verbalizations will be recorded by the videocamera along with an image of your work area during parts 1 and 3 of the experiment.

## Studying the Program (Reuse Orientation)

Please read and attempt to comprehend the program which you will be given **with the objective of later reusing** it in the solution of another problem.

You will be given a hardcopy of the program and also an online version accessible by the editor. The hardcopy contains the files in alphabetical order, and you can change the order or staple the pages if you want.

You may take notes during the study of the program. If you would like to take such notes either on the program listing or on separate sheets of paper, you may use the pens provided by the experimenter. From time to time the experimenter will ask you to change the color of pen you are using.

You may also use the textbook provided to look up information about C++ during the experiment.

During this study phase, you may view the program using the editor and execute it, but you may not change it. You will reutilize the program in a later phase of the experiment.

While you study the program, please **think-aloud**.

## Studying the Program (Documentation orientation)

Please read and attempt to comprehend the program which you will be given **with the objective of later writing documentation** for it.

You will be given a hardcopy of the program and also an online version accessible by the editor. The hardcopy contains the files in alphabetical order, and you can change the order or staple the pages if you want.

You may take notes during the study of the program. If you would like to take such notes either on the program listing or on separate sheets of paper, you may use the pens provided by the experimenter. From time to time the experimenter will ask you to change the color of pen you are using.

You may also use the textbook provided to look up information about C++ during the experiment.

During this study phase, you may view the program using the editor and execute it, but you may not change it. You will document the program in a later phase of the experiment.

While you study the program, please **think-aloud**.

## Instructions for Answering Comprehension Questions

This part of the experiment consists of 54 questions about the program.

A question will appear on the screen and you should answer yes or no as quickly as possible without making too many errors.

Keep your finger on the response button so that you can make your choice quickly. You will have a maximum of 30 seconds to answer each question. After you respond to a question there will be a short pause before the next question appears. Please stay alert so that you begin reading each question immediately when it appears. You will be asked if you want to take a break after each 18 questions.

You do not need to think-aloud while answering the questions, and you will not be videotaped.

We will begin with 4 very simple practice questions to familiarize you with the procedure, the 12 questions to determine your reading speed.

After these warm-up questions, questions of different types will be asked. These questions types are described on the following page.

## Types of Questions

Does the program the define class "X"?
 [X is the name of a class]

Does class "X" inherit from class "Z"?
 [X and Z are the names of classes; Inheritance may be direct or through intermediary classes]

Is the field "X::y" of the type "Z"?
 [You should answer yes if field y of class X is of type Z or if this field is a pointer to an instance of of the class Z]

Does the program allow you to do A?
 [A is is the description of one or several actions]

Does class "X" call a member function of class "Z"?
 [You should answer yes if class X calls class Z directly or if it communicates indirectly through a series of successive calls or dynamic binding]

Does the "X::y" do A?
 [X is the name of a class, y a function in the class, and A the description of one or several actions]

In "X::y" does the value of "v1" affect the value of "v2"?

[where X is the name of a class, y is a function in the class, and v1 and v2 are two variables. You should answer yes if v1 can affect the value of v2 through direct assignment, through controlling a conditional, or through parameter passing]

In "X::y" is A done before B?
[where X is the name of a class, y is the name of a function in the class, and A and B are the descriptions of one or several actions]

Does the program contain the code fragment: I
[I is a fragment.of C++ code]

### Instructions for the Reuse Task

You will next be given the statement of a problem for which you are to design and implement a solution in C++. To solve the problem, we suggest that you **make use of the university scheduling program** which you studied previously.

To do this task, you may make use of the hardcopy and the online version. You may use the editor, compiler-linker, and debugger, and you may also take notes with paper and pen and refer to the textbook.

You will have a total of 1 and 1/2 hours to solve the problem, and you should make as much progress as you can in designing and implementing a solution during that time.

Please **think-aloud** as you work. In your verbalizations please indicate what concepts or parts of the program you are reusing and why.

### Instructions for the Documentation Task

You will next be asked to document the **university scheduling program** which you previously studied. Your objective should be to document the code for another programmer who may later have to work with the program to make modifications or extensions.

You may make use of the hardcopy and the online version, but you should place your documentation in the the online version. However, if you wish to provide documentation which is difficult to put in the online file, you may write it on paper or the listing of the program, making explicit to the experimenter that it is a part of the documentation and not just notes. To do this task, you may use the editor, compiler-linker, and debugger, and you may also take notes with paper and pen and refer to the text book.

You will have a total of 1 and 1/2 hours to document the program, and you should make as much progress as you can during that time.

Please **think-aloud** as you work.

## APPENDIX 4
## Timing of the experiment

| | |
|---|---|
| Introduction | 5 minutes |
| Study program | 35 minutes |
| 1st questions set | 45 minutes |
| BREAK | 10 minutes |
| Reuse/ documentation task | 90 minutes |
| 2nd question set | 45 minutes |
| Background questionnaire | 15 minutes |
| | __________ |
| | 245 minutes = 4 hours 5 minutes |